# Distributed Data Processing Frameworks for Big Graph Data

Afsin Akdogan, Hien To

*University of Southern California*

*Los Angeles, CA, 90089, USA*

`[aakdogan,hto]@usc.edu`

## I. INTRODUCTION

Recently we create so much data (2.5 quintillion bytes every day) that 90% of the data in the world today has been created in the last two years alone [1]. This data comes from sensors used to gather traffic or climate information, posts to social media sites, photos, videos, emails, purchase transaction records, call logs of cellular networks, etc. This data is *big data*. In this report, we first briefly discuss what programming models are used for big data processing, and focus on graph data and do a survey study about what programming models/frameworks are used to solve graph problems at very large-scale.

In section 2, we introduce the programming models which are not specifically designed to handle graph data but we include them in this survey because we believe these are important frameworks and/or there have been studies to customize them for more efficient graph processing. In section 3, we discuss some techniques that yield up to 1340 times speedup for some certain graph problems when applied to Hadoop. In section 4, we discuss vertex-based programming model which is simply designed to process large graphs and the frameworks adapting it. In section 5, we implement two of the fundamental graph algorithms (Page Rank and Weight Bipartite Matching), and run them on a single node as the baseline approach to see how fast they are for large datasets and whether it is worth to partition them.

## II. BIG DATA PROCESSING FRAMEWORKS

Distributed data processing models has been one of the active areas in recent database research. Several frameworks have been proposed in database literature. Figure 1 shows the release date of some of the successful frameworks. The arrows show the dependencies among the models. For example, Hive converts the scripts written with its own language into MapReduce tasks so there is an arrow connecting them.

In 2004 Google proposed the MapReduce functional programming model which provides regular programmers the ability to produce parallel distributed programs easily. Although it is extremely scalable, this simplified framework is either not capable of modelling and solving many of the problems or it is very inefficient to model every single problem with this **general-purpose** framework. For example, it has been shown that using Hadoop on relational data is at least a factor of **50** less efficient than it needs to be [10, 11]. To address this inefficiency, **problem** and **data-specific** programming models have been developed such as Facebook's Hive [3] for SQL-like workloads, Yahoo's Pig Latin [4] for iterative data processing, Google's Pregel [5] for graph processing, Microsoft's Dryad [8], and SpatialHadoop [12] for geospatial data processing, etc. In this section we discuss these programming models.

### A. MapReduce

With ever-increasing popularity of mobile devices, social media and web-based services such as emails, we create so much data that the on-hand relational database tools are not capable of handling and processing the data at this scale. Therefore, in 2004 Google proposed the MapReduce functional programming model which provides regular programmers the ability to produce parallel distributed programs easily, by requiring them to write only the simple map and reduce functions. Figure 2 shows how Hadoop [9] (an open-source implementation of MapReduce model) processes the data of four split in parallel where there are three map machines and two reduce machines.

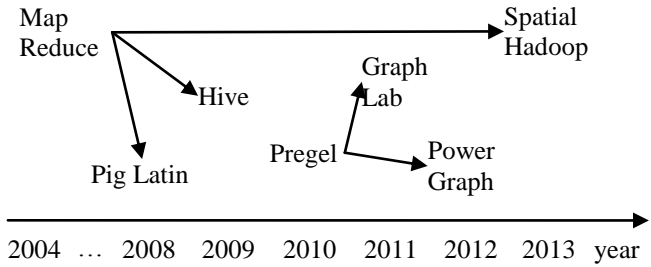

Figure 1: Timeline for distributed programming models for big data processing

The problem with Hadoop is that its strength is also its weakness. Hadoop gives the user power to scale different data management problems. However, this flexibility that allows the user to perform inefficient operations and not care because they can add more computing nodes and use Hadoop's scalability to hide inefficiency in user code, and since it is designed for batch data processing, they can let their process run in the background and not care about how long it will take for it to return. For example, it has been shown that using Hadoop on relational data is at least a factor of **50** less efficient than it needs to be [10, 11]. Instead of having a single general-purpose programming model, it is more efficient to group certain types of problems and come up with a framework works only for that domain. In the following sections we briefly introduce these frameworks.

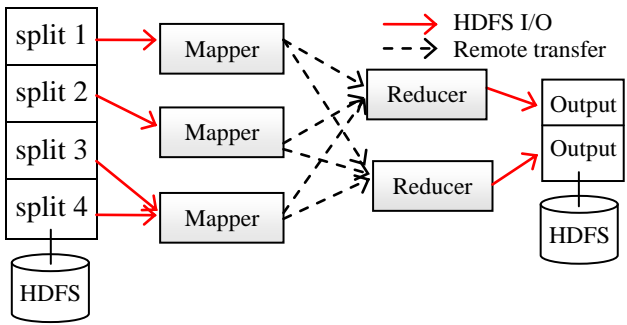

Figure 2: The data flow in Hadoop Architecture

### B. Hive

Hive is an open-source framework built on top of Hadoop. Hive supports queries expressed in a SQL-like declarative language - HiveQL, which are first translated into MapReduce tasks and then executed on Hadoop [3]. In addition, HiveQL supports custom map-reduce scripts to be plugged into queries. The language includes a type system with support for tables containing primitive types, collections like arrays and maps, and nested compositions of the same. The underlying IO libraries can be extended to query data in custom formats. In Facebook, the Hive warehouse contains several thousand tables with over 700 terabytes of data and is being used extensively for both reporting and ad-hoc analyses by more than 100 users [3].

### C. Pig Latin

The fact that MapReduce programming model is too low-level leads to a tremendous amount of custom user code that is hard to maintain and reuse. Pig Latin addresses these problems by defining a higher-level new language which is a combination of SQL and the low-level, procedural style of map-reduce. Pig is fully implemented at Yahoo! and it dramatically reduces the time required for the development and execution of their data analysis tasks, compared to using Hadoop directly [4]. The main difference between Pig Latin and Hive is that Pig is more declarative such that programmers can use variables, loops and if/else clauses. However, the underlying structure is still the same as they both translate the scripts written in higher-level languages into MapReduce jobs.

### D. Spatial Hadoop

SpatialHadoop [12] is a MapReduce framework designed specifically to work with spatial data. It adapts some of the well-known spatial index structures such as R-tree for efficient processing of queries such as Nearest Neighbor and its variations. Since we focus on distributed programming models for graph data processing, we will not get into details about this framework.

## III. TECHNIQUES TO CUSTOMIZE HADOOP FOR MORE EFFICIENT GRAPH PROCESSING

There has been tremendous amount of effort to optimize Hadoop's performance. One of the significant research studies to customize Hadoop for more efficient graph processing is [18, 19]. In this paper, authors achieve tremendous amount of speedup, up to 1340 times for sub-graph pattern matching queries, in the following three ways:

1. ***Partitioning***: Hadoop hash partitions data across data nodes. This results in data for each vertex in the graph being randomly scattered across the cluster. Therefore, data elements (vertices) that are close to each other in the graph might end up in different nodes. This type of partitioning is quite suboptimal for graph processing. Rather than hash partitioning, using a clustering algorithm to graph partition data across nodes in the Hadoop cluster improves the performance.

2. ***Replication***: Hadoop has a very simple replication method, where all data is replicated a fixed number of times, by default 3, across the cluster. When it comes to replication it is inefficient to treat all of the data equally because the data on the border of any particular partition is more important to replicate than the data that is internal to a partition. This is because vertexes that are on the border of a partition might have several of their neighbours stored on different computing node. For the same reasons why it is a good idea to graph partition data to keep graph neighbours local, it is a good idea to replicate data on the edges of partitions so that vertexes are stored on the same computing node as their neighbours.

3. ***Storage***: Hadoop stores the data on a distributed file system (HDFS) and HDFS is not optimized for graph data. By replacing the physical storage system with graph-optimized storage, but keeping the rest of the system as the same, it is possible to increase the efficiency of the system.

Each of the above improvements speeds up the system by a factor of **10** [18].

## IV. VERTEX-BASED GRAPH DATA PRPOGRAMMING MODELS

### A. *Static versus Dynamic Graphs*

We can simply classify graphs as *static* and *dynamic* graphs based the update frequency of the dataset. For example, the graph datasets in communication networks and VLSI designs are static graphs since they are subject to infrequent insert/delete edges or vertices. For this kind of workloads, which have low-frequent updates, the conventional way of doing graph operations is first to pre-compute and run the desired algorithm on the precomputed data. For example, in [3] the goal is to find shortest path for a given source and a destination vertex. Instead of computing shortest path for every query in real-time, the algorithm pre-computes and stores all possible combinations of shortest paths and stores them efficiently, later to be used as an index structure in the query processing time. The main problem with the pre-computation approach is that it cannot handle frequent updates and the cost of keeping the pre-computed index might be even more costly than re-creating the index structure from scratch. This update problem is an impediment to the applicability of pre-computation techniques to the large *dynamic graphs* such Web and social network graphs especially for large datasets. The better way of processing large *dynamic graphs* is to utilize distributed data processing models.

### B. *Vertex-based Model*

There are several programming models to express distributed computations on large datasets such as MapReduce [2]. This general-purpose model is sometimes used to mine *large dynamic graphs* [13, 14, 18]; however, this leads to suboptimal performance. For instance, expressing a graph algorithm as a sequence of MapReduce statements requires passing the entire state of the graph between map and reduce steps. This limitation of the model incurs tremendous amount of unnecessary network I/O resulting in poor performance. Graph data processing is obviously not the strong suit of these multipurpose models. Therefore, a new model, vertex-based, which is particularly designed to mine wide range of very large graphs with billions of vertices and trillions of edges, has been developed. In this section we will introduce Pregel [5] and its following variations GraphLab [6] and PowerGraph [7].

### C. Pregel

The high-level organization of Pregel programs is inspired by Bulk Synchronous Parallel model [15]. Pregel computations are composed of a sequence of iterations, called *supersteps*. During a *superstep* the framework calls a user-defined function for each vertex in parallel. The function defines behaviour at a single vertex V and a single superstep S. It can read messages sent to V in superstep S - 1, send messages to other vertices that will be received at superstep S + 1, and modify the state of V and its outgoing edges. A message can be sent to any vertex whose identity is known. Obviously in Pregel model we keep the vertices and edges in the servers and use network only to pass messages among vertices. A vertex can either be active or inactive and can change its state (see Figure 3). The only difference between these two states is that active vertices can send messages and inactive ones cannot.

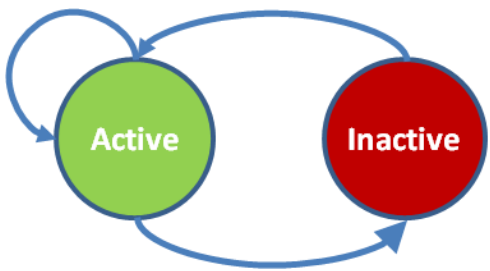

Figure 3: Vertex-based programming model

One of the key challenges in designing scalable graph algorithms with Pregel model is to find the most appropriate partitioning technique for the given problem. Due to the changing degree of parallelism over the course of execution, some of the servers might stay idle for a while. For example, initially there are two computing nodes *A* and *B* with 100 active vertices each. At some point of the algorithm, there might be a case that node *A* still has 100 active vertices and *B* only has 10. That means node *A* does 10 times more work than node *B*. This unbalanced workload across servers would decrease the level of parallelism and worsens *load-balancing*. To overcome this problem, partitioning method should be selected carefully.

### D. GraphLab

sssss

### E. PowerGraph

wwww

## V. PERFORMANCE EVALUATION

In this section we implemented some of the fundamental graph algorithms such as Page Rank and Weight Bipartite Matching. Our goal is to figure out if it is worth to partition and parallelize these problems. Therefore, as a baseline approach we run the experiments on a *single computing node* with 4GB RAM and Intel core 2 duo 3.16GHz CPU and we use only a *single thread*.

### A. Page Rank

Page Rank is a link analysis algorithm used by the Google web search engine which assigns a weight to each element of a hyperlinked set of documents. Page Rank is an iterative algorithm and typically it needs to run for at least three and at most 100 iterations to compute the rankings for the given web graph. The time complexity of the algorithms is:

$$O(|E| \cdot I)$$

Where *E* is the number of edges and *I* is the number of iterations. We run our experiments with varying number of edges, *I* is fixed and always 100. We exclude the data loading time from disk. *The entire graph fits in memory* and the reported numbers only are only for solving the given problem. The figure below shows that the total execution time *linearly increases* as a function of the number of edges. The main use of Page Rank algorithm is to rank pages on the Web. We don't have the entire web graph; however, we can estimate how long it would take to compute page rank for the Web using the values below. According to [16], there are 14 billion pages and approximately each page is linked to 7 other pages on average on the Web. If it takes around 471 minutes to calculate page rank for 100 million edges, it would take $((14 \cdot 10^9 \cdot 7) / 10^8) \cdot 471$ *minutes* $\approx 320$ days. It is obvious that we need to utilize distributed programming models to solve page rank for very large datasets.

It is also important to note that we use a single thread on a single node and the amount of sequential code is very close to %0 for Page Rank as essentially every data element can be processed independently. Therefore, it can speed up almost linearly as more memory and cores are used (***Amdahl's law***) [17].

| Edges | Disk Loading (minute) | PageRank (minute) |
|-------|-----------------------|-------------------|
| 1M | 0.12 | 4.5 |
| 10M | 1.25 | 49 |
| 100M | 14 | 471 |

Figure 4: Total runtime of Page Rank (*I*=100) on a single computing node for varying number of edges of Orkut.com dataset and disk loading time for corresponding dataset

### B. Weight Bipartite Matching

The input to a bipartite matching algorithm consists of two distinct sets of vertices with edges only between the sets. The desired output is a subset of edges with no common endpoints. This problem is one of the fundamental graph problems and it is also implemented and presented in Pregel paper along with Page Rank. There are several algorithms have been developed to process bipartite matching. Time and space complexity of the one we use in the experiment are:

time complexity: $O(|V|^2)$

space complexity: $O(|V|^2)$

Where V is the number of vertices. We run our experiments for small datasets since we run out of memory after 10,000 vertices. Figure shows the time total execution time for varying number of vertices. Considering the memory requirement and time complexity of the problem, we conclude that weight bipartite matching problem should be partitioned and parallelized especially for large datasets.

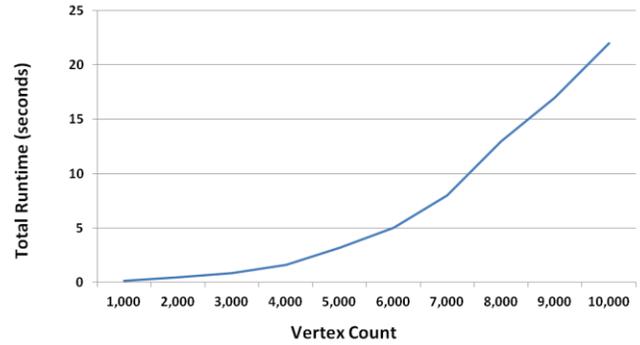

Figure 5: Total runtime of Weight Bipartite Matching Problem on a single computing node for varying number of vertices.

### VI. CONCLUSIONS